\documentclass[preprintnumbers,showpacs,amsmath,amssymb,floatfix,9pt,prd,onecolumn,
superscriptaddress,nofootinbib]{revtex4}
\usepackage{latexsym}
\usepackage{epsfig}
\usepackage{amssymb}
\begin{document}

\title{Stability of Thin-Shell Wormholes from Noncommutative BTZ Black Hole}

\author{Piyali Bhar}
 \email{piyalibhar90@gmail.com}
\affiliation {Department of Mathematics, Jadavpur University,
 Kolkata-700032, India}

\author{Ayan Banerjee}
\email{ayan\_7575@yahoo.co.in}
\affiliation {Department of Mathematics, Jadavpur University,
 Kolkata-700032, India}

\begin{abstract}
In this paper, we construct thin-shell wormholes in (2+1)-dimensions
from noncommutative BTZ black hole by applying the cut-and-paste procedure
implemented by Visser. We calculate the surface stresses localized at the wormhole
throat by using the Darmois-Israel formalism, and we find
that the wormholes are supported by matter violating the energy conditions.
In order to explore the dynamical analysis of the wormhole throat, we consider
that the matter at the shell is supported by dark energy equation of state
$ p=\omega\rho$ with $\omega< 0$. The stability analysis is carried out of these  wormholes
to linearized spherically symmetric perturbations around static solutions.
Preserving the symmetry we also consider the linearized radial perturbation
around static solution to investigate the stability of wormholes  which
explored  by the parameter $\beta$ (speed of sound).

\end{abstract}

\maketitle

\section{Introduction}
Wormholes and thin-shell wormholes are very interesting topic to the
researcher for last two decades. Historically the concept of wormholes was first
suggested by Flamm \cite{Flamm} by means of the standard embedding diagram. After nineteen
years a similar construction (wormhole type solution) attempted by Einstein's and
Rosen (1935) \cite{Einstein and Rosen} known as `Einstein-Rosen Bridge'. They tried to investigate the
fundamental particles like electrons as space-tunnels threaten by electric lines of
force rather than to promote inter-universe travel. But these types of wormholes were not
traversable. Interest in traversable wormholes, as a hypothetical shortcuts in space-time
after introduced by Morris and Thorne \cite{MT1988}. Such types of wormholes are both traversable
and stable as a solution of Einstein's field equations having two asymptotically flat  regions
(same universe or may be two separate universe) connected by
a minimal surface area, called throat, satisfying  the flare-out condition \cite{HV1997}. Traversable
wormholes have some issues such as the violations of energy conditions \cite{Visser 1995,Visser 2002},
the mechanical stability etc. which also stimulate researcher in several branches.

 Though, it is very difficult to deal with the exotic matter
(matter not fulfilling the energy conditions), Poisson and Visser \cite{Poisson} constructed
a thin-shell wormhole  by cutting and pasting
two identical copies of the Schwarzschild solution at $a > 2M$, to form a geodesically
 complete new one with a shell placed in the joining surface.
This construction restricts exotic matter to be placed at the wormhole throat and
consider linearized radial perturbation around a static solution, in the spirit of \cite{Brady,Balbinot}
for the stability of wormholes. Though the region of stability lies in a unexpected
patch due to the lack of knowledge of known  equations of state of exotic matter. Therefore,
it is perhaps important to investigate the stability of static wormhole solutions
by using specific equation of state (EoS) or by considering a linearized radial perturbations
around a static solution. The choice of equation of state for the description of matter
violating energy conditions present in the wormhole throat
has a great relevance in the existence and stability of wormhole
static solutions. Several models for the matter leading to such situation have been
proposed \cite{Sahni,Peebles,Padmanabhan}: one of them is the `dark energy' models
are parameterized by an equation of state w = p/$\rho$ where p is the spatially homogeneous
pressure and $\rho$ is the dark energy density. A specific form of
dark energy, denoted phantom energy, possessing with the peculiar property of
w $<$ -1. As the phantom energy equation of state, p = w$\rho$ with
$w < -1$, is now a fundamental importance to investigate
the stability of these phantom thin-shell wormholes have been studied
in \cite{Jamil,Lobo2005,Lobo2006,Sushkov,Kuhfittig,Faraoni,Rahaman2006(38),Bronnikov,Bronnikov2007}.
Another one is generalized Chaplygin gas, with equation of state $p\rho^{\alpha}=-A$
($0<\alpha\leq 1$), also denoted as quartessence, based on a negative pressure fluid,
has been proposed by Lobo in Ref. \cite{Lobo2006(73)}. The existence of  GCG model
remains hypothetical but the astrophysical observations, supernovae
data \cite{Bento,Bertolami}, cosmic microwave background radiation \cite{Bento2003,Bento2003(575),Carturan,Bento2003(35)},
gravitational lensing \cite{Silva,Dev,Fabris}, gamma-ray bursts \cite{Silva2006}, have suggested the
presence of phantom energy in our observable universe. A great amount of works have been found
in studying the Chaplygin thin-shell wormholes in \cite{Eiroa2011,Eiroa2009,Eiroa2007,Sharif2013,Sharif2013(1305),Eiroa2012,Jamil2009,Gorini}.
The study of thin-shell wormholes have been extended
include charge and cosmological constant, as well as other features (see, for example,
references \cite{Eiroa2004,Lobo2004(391),Thibeault,Eiroa2005(71),Rahaman2006,Richarte,Dotti,Usmani,
Rahaman2010,Lemos2004,Lemos2008,Rahaman2009}).

Recent, years among different outcomes of string theory, we focus our
study on noncommutative geometry which is expected to be relevant at the Planck scale
where it is known that usual semiclassical considerations break down.
The approach is based on the realization that coordinates (which assumes extra dimensions)
may become noncommuting operators in a D-brane \cite{Smailagic,Nicolini,Spallucci,Nicolini2010}.
The noncommutativity of spacetime can be encoded in the commutator [x$^\mu$, x$^\nu$] = i $\theta^{\mu\nu}$,
where $\theta^{\mu\nu}$ is an anti-symmetric matrix, similar to the way that the Planck constant
 $\hslash$ discretizes phase space \cite{Gruppuso}. Noncommutative geometry is an
intrinsic property of space-time which does not depend on particular features such as curvature,
rather it is a point-like structures by smeared objects, a minimul length responsible for
delocalization of any point like object \cite{Smailagic(36)}. A number of studies
on the exact wormhole solutions in the context of noncommutative geometry can be found in
the literature \cite{Garattini,Garattini2013,Peter,Rahaman1305,Ayan}. Recently,
Rahaman et al.\cite{Rahaman87} have investigates the properties of a BTZ black
hole \cite{BTZ1992} constructed from the exact solution of the Einstein field equations in a
(2 + 1)-dimensional anti-de Sitter space-time in the context of noncommutative geometry. They showed
that noncommutative geometry background is able to account for producing stable circular orbits,
as well as attractive gravity, without any need for exotic dark matter. Kuhfittig \cite{kh} showed that a special class of thin shell wormholes could be possible that are unstable in classical general relativity but are stable in a small region in noncommutative spacetime. Bhar and Rahaman \cite{pb} have observed that the wormhole solutions exist only in four and five dimensions ; however,in higher than five dimensions no wormhole exists. For five dimensional spacetime, they got a wormhole for a restricted region. In the usual four dimensional spacetime, they obtain a stable wormhole which is asymptotically flat.  Garattini and Lobo \cite{lobo}
obtained a self-sustained wormhole in noncommutative geometry. On the other hand, pure gravity
in (2+1)-dimensions is not as trivial as it seems at first sight. The specific properties of (2 + 1)-
dimensional space-time allow us to improve our understanding of the classical physics it defines before
tackling the (3+1) problem. The objective of this paper is to construct thin-shell wormholes from
noncommutative BTZ black hole in (2+1)-dimensional gravity.
Inspired by this work we provide thin-shell wormholes from noncommutative BTZ black
hole in (2+1)-dimensional gravity. A well studied of thin-shell wormholes in lower
dimensional gravity have been found in \cite{Ayan2013,Eiroa2013,Mazharimousavi2014,Eiroa2014,Ayan2012}.

The objective of this paper is to construct thin-shell
wormholes from noncommutative BTZ black hole in (2+1)-dimensional
gravity. We discuss various properties of wormholes and investigate its stability regions
under linearized radial perturbation. This paper is organized as follows: In
section II, wormholes are constructed by cutting
and pasting geometries associated to the noncommutative BTZ black
holes by applying the Darmois–Israel formalism. In section III,
we discuss effect of gravitational field and calculate
the total amount of exotic matter in section IV. In section V, we  also consider
the specific case of static wormhole solutions, and consider several equations
of state. In section VI,  stability analysis is carried out
for the dynamic case by taking into account specific surface
equations of state, and the linearized stability analysis
around static solutions is also further explored. Finally, section VII, discusses the results of the paper.

\section{Construction of Thin-shell Wormhole}

Recently Rahaman {\em et al.} \cite{Rahaman87} have found a  BTZ black hole inspired by
noncommutative geometry with  negative cosmological
constant $\Lambda$. The metric is given by
\begin{equation}
ds^{2}=-f(r)dt^{2}+[f(r)]^{-1}dr^{2}+r^{2}d\phi^{2},
\end{equation}
where
\begin{equation}
f(r)=M\left[2e^{-\frac{r^{2}}{4\theta}}-1\right]-\Lambda r^{2}.
\end{equation}
Here, M is the total mass of the source. Due to the uncertainty,
it is diffused throughout a region of linear dimension $\sqrt{\theta}$.
In the limit, $\frac{r}{\sqrt{\theta}}\rightarrow \infty$, the Eq. (1) reduces to a
BTZ black hole. To construct thin-shell wormhole  we take two identical copies
from noncommutative BTZ black hole with
$r\geq a$:
\[ \mathcal{M}^\pm = ( x \mid r \geq a ),  \]
and stick them together at the  junction surface
\[ \Sigma = \Sigma^\pm = ( x \mid r = a ),  \]

to get a new geodesically complete manifold
$ \mathcal{M} = \mathcal{M}^+
+\mathcal{M}^- $. The minimal surface area
$\Sigma$, referred as a throat of wormhole where we
take $ a > r_h $, to avoid the event horizon.
Thus a single manifold $\mathcal{M}$,  which we obtain
connects two non-asymptotically flat regions at a junction surface, $\Sigma$,
where the throat is located. The wormhole throat  $\Sigma$, is a timelike hypersurface
define by the parametric equation of the form $f\left(x^{\mu}(\xi^i)\right)=0$,
where we define coordinates $\xi^{i}= \left(\tau, \phi\right)$, with $\tau$
is the proper time on the hypersurface. The induced metric on $\Sigma $ is given by
\begin{equation}
ds^{2}=-d\tau^{2}+a^{2}d\phi^{2}.
\end{equation}
In our case the junction surface is a one-dimensional ring of matter.
To understand the stability analysis under perturbations preserving the symmetry, we assume
the radius of the throat to be a function of proper time i.e., a = a($\tau$ ).
Further we assume that the geometry remains static outside the
throat, regardless that the throat radius can vary with time, so no gravitational waves are present.

 We shall use the Darmois-Israel formalism to
determine the surface stresses at the
junction boundary \cite{Israel,Papapetrou}. The intrinsic surface stress-energy tensor,
$S_{ij}$ , is given by the Lanczos equations in the form
\begin{equation}
S^{i}_j=-\frac{1}{8\pi}\left(\kappa^{i}_j-\delta^{i}_j\kappa^{l}_l \right),
\end{equation}
where  $K_{ij}$ is not continuous for the thin shell across $\Sigma$, so that for
notational convenience, the discontinuity in the second fundamental form is defined as
$\kappa_{ij}=K_{ij}^{+}-K_{ij}^{-}$, with

\begin{equation}K^{i\pm}_j =  \frac{1}{2} g^{ik}
\frac{\partial g_{kj}}{\partial \eta}   \Bigl\lvert_{\eta =\pm 0} =
\frac{1}{2}   \frac{\partial r}{\partial \eta} \Bigl\lvert_{r=a}
 ~ g^{ik} \frac{\partial g_{kj}}{\partial r}\Bigl\rvert_{r=a},
\end{equation}
where $\eta$ is the Riemann normal coordinate at the junction which has positive signature in the manifold described by exterior space-time and negative signature in the manifold described by interior space-time.
Now using the symmetry property of the solution, $K^{i}_j$, can be written as
\begin{equation}
K^{i}_j=\left(
  \begin{array}{cc}
    \kappa^{\tau}_{\tau} & 0 \\
    0 & \kappa^{\phi}_{\phi} \\
  \end{array}
\right).
\end{equation}
Thus, the surface stress-energy tensor may be written in terms of the surface energy
density, $\sigma$, and the surface pressure, p, as $S^{i}_j=\text{diag}\left(-\sigma, p, p\right)$,
which taking into account the Lanczos equations, reduce to
\begin{eqnarray} \sigma &=&  -\frac{1}{8\pi}  \kappa _\phi^\phi,\\
\label{eq38} v &=&  -\frac{1}{8\pi}  \kappa _\tau^\tau.
\label{eq39}
 \end{eqnarray}
This simplifies the determination of the surface stressenergy
tensor to that of the calculation of the non-trivial
components of the extrinsic curvature, or the second fundamental
form. After some algebraic manipulation, we obtain that the
energy density and the tangential pressures can be recast
as
\begin{equation}
\sigma=-\frac{1}{4\pi a}\sqrt{f+\dot{a}^{2}},
\end{equation}
\begin{equation}
p=\frac{1}{8\pi}\frac{2\ddot{a}+f'}{\sqrt{f+\dot{a}}^{2}},
\end{equation}
respectively. Here $p $ and $\sigma$ obey the conservation equation
\begin{equation}
\frac{d}{d\tau}(\sigma a)+p\frac{d}{d\tau}(a)=0,
\end{equation}
or
\begin{equation}
\dot{\sigma}+\frac{\dot{a}}{a}(p+\sigma)=0,
\end{equation}
where $\dot{a}\equiv\frac{da}{d\tau}$ and $f'\equiv\frac{df}{da}$, respectively. For a
static configuration of radius a, we obtain (assuming
$\dot{a} = 0$ and $\ddot{a} = 0$) from Eqs. (9) and (10)
\begin{equation}
\sigma_0=-\frac{1}{4\pi a}\sqrt{M \left(2e^{-\frac{a^{2}}{4\theta}}-1\right)-\Lambda a^{2}},
\end{equation}
\begin{equation}
p_0=-v_0=\frac{1}{4\pi}\left[ \frac{-\Lambda a-\frac{Ma}{2\theta}e^{-\frac{a^{2}}{4\theta}}}{\sqrt{M \left(2e^{-\frac{a^{2}}{4\theta}}-1\right)-\Lambda a^{2}}}\right].
\end{equation}
The energy condition demands, if $\sigma \geq 0$ and $ \sigma+ p\geq 0$ are
satisfied, then the weak energy condition (WEC) holds and by continuity, if $ \sigma+ p\geq 0$ is
satisfied, then the null energy condition (NEC) holds. Moreover, the strong energy(SEC) holds,
if $ \sigma+ p\geq 0$ and $ \sigma+ 2p\geq 0$ are satisfied. We get from
 Eqs. (13) and (14), that $\sigma< 0$ but $ \sigma+ p\geq 0$ and
 $ \sigma+ 2p\geq 0$, for all values of M and $\theta$, which show that the
shell contains matter, violates the weak energy condition
and obeys the null and strong energy conditions which is shown in Fig. \textbf{(4)} (left).

Using different values of mass (M) and noncommutative parameter ($\theta$), we
plot $\sigma$ and p as a function of `a', shown in Figs. \textbf{1-3}.

\begin{figure*}[thbp]
\begin{tabular}{rl}
\includegraphics[width=9cm]{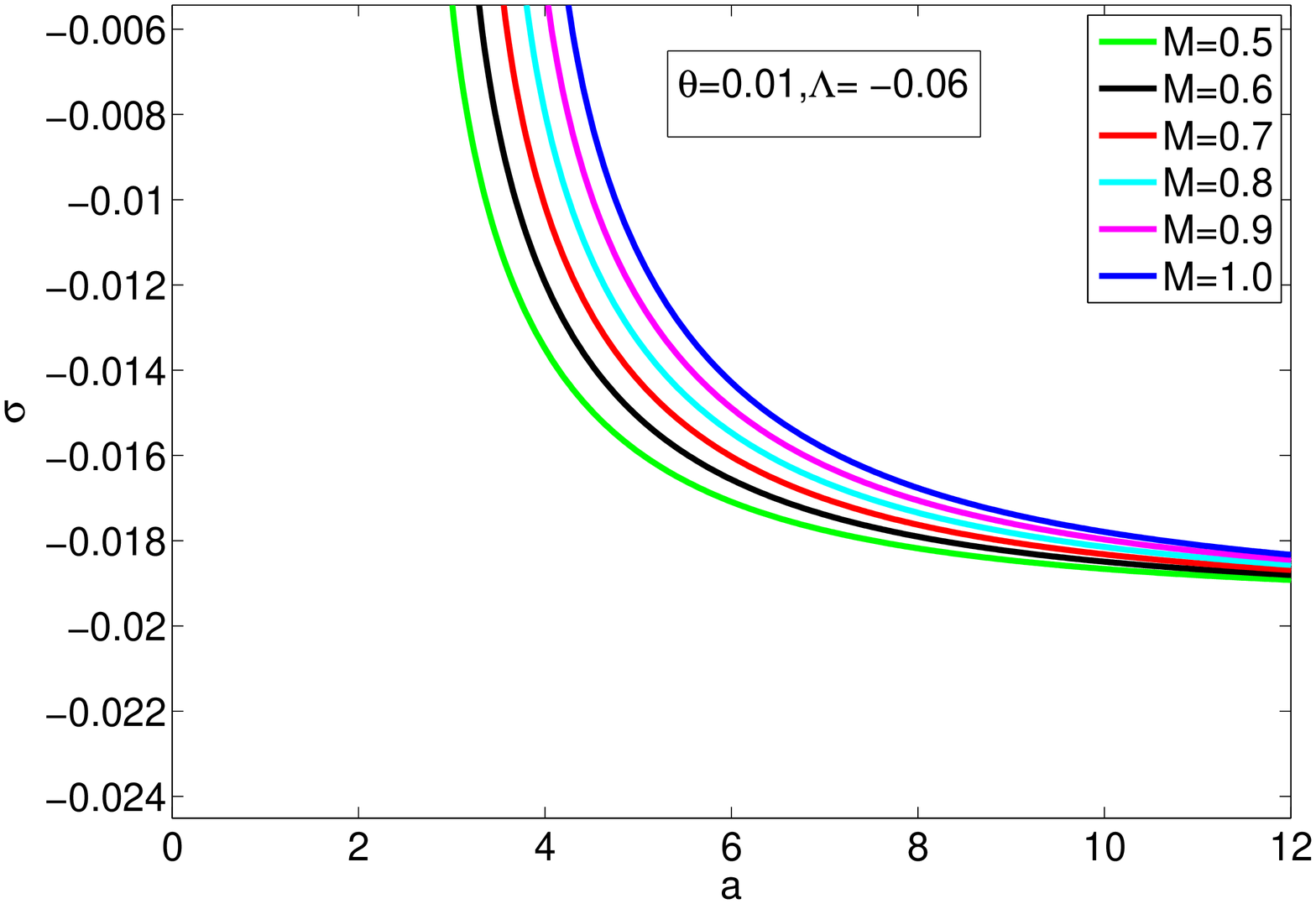}&
\includegraphics[width=9cm]{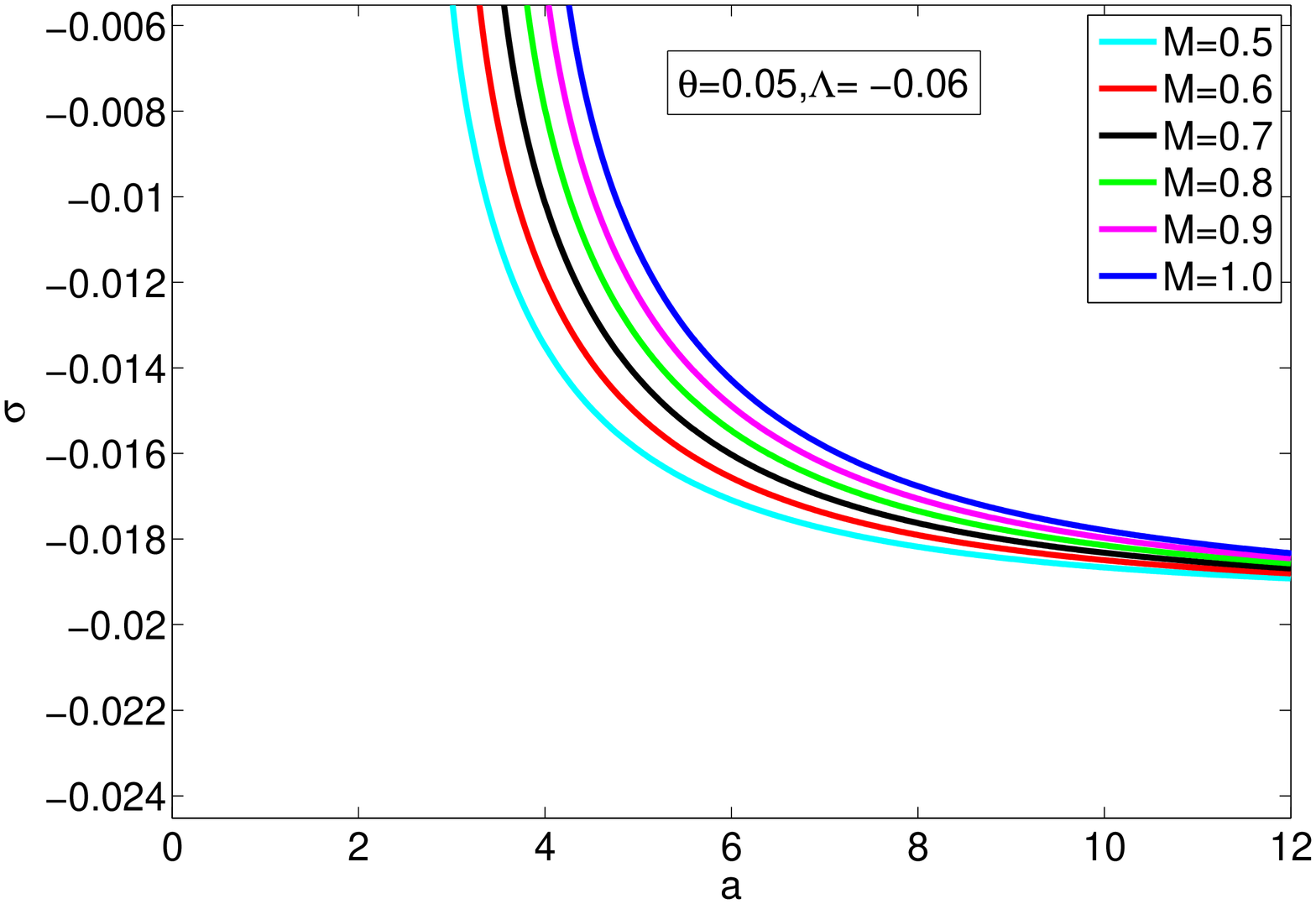} \\
\end{tabular}
\caption{Surface Energy density $\sigma$ has been plotted against $a$ for different values of mass $M$
when noncommutative parameter $\theta=0.01$ (left) and for  $\theta=0.05$ (right).}
\end{figure*}

\begin{figure*}[thbp]
\begin{tabular}{rl}
\includegraphics[width=9cm]{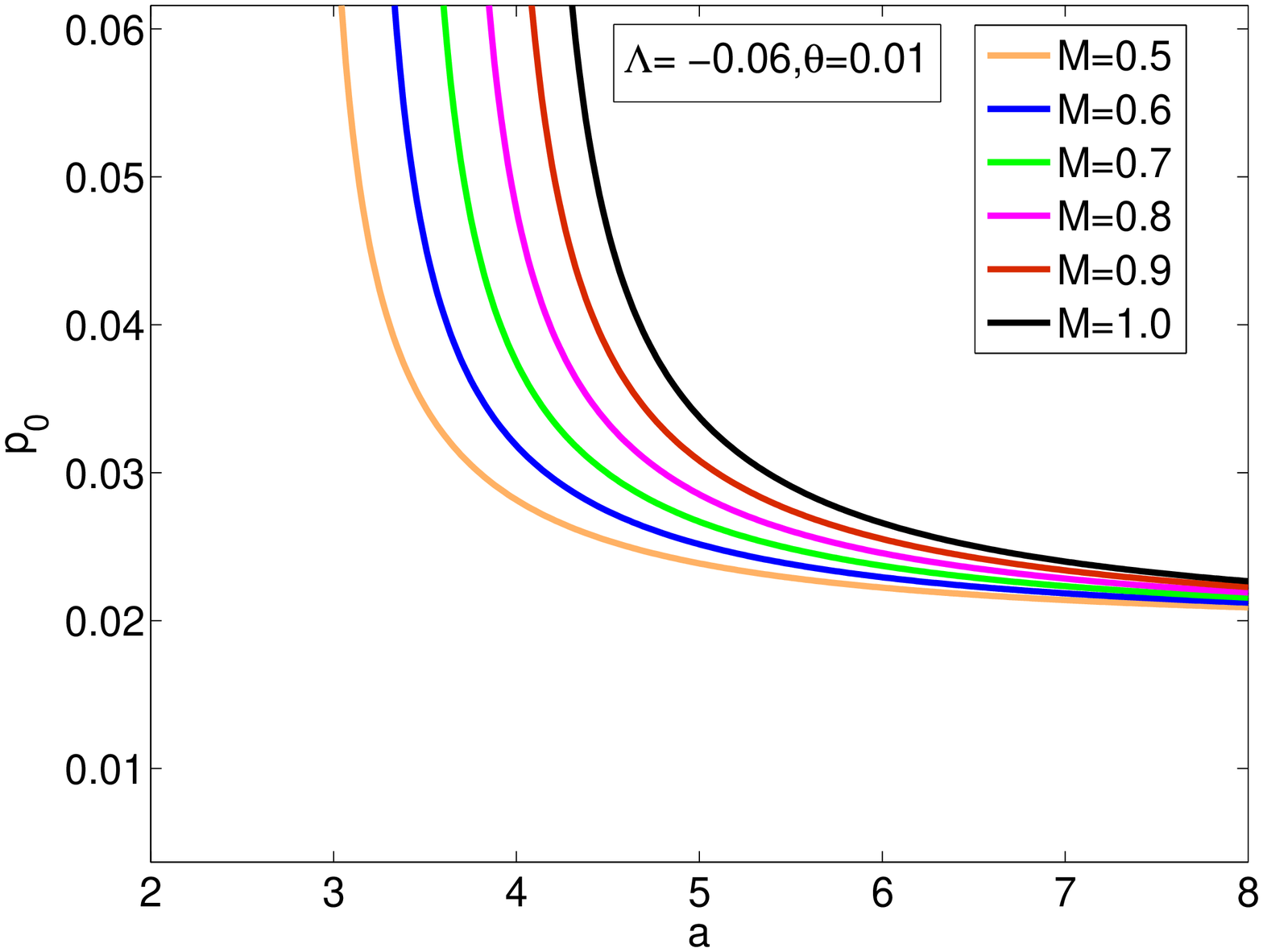}&
\includegraphics[width=9cm]{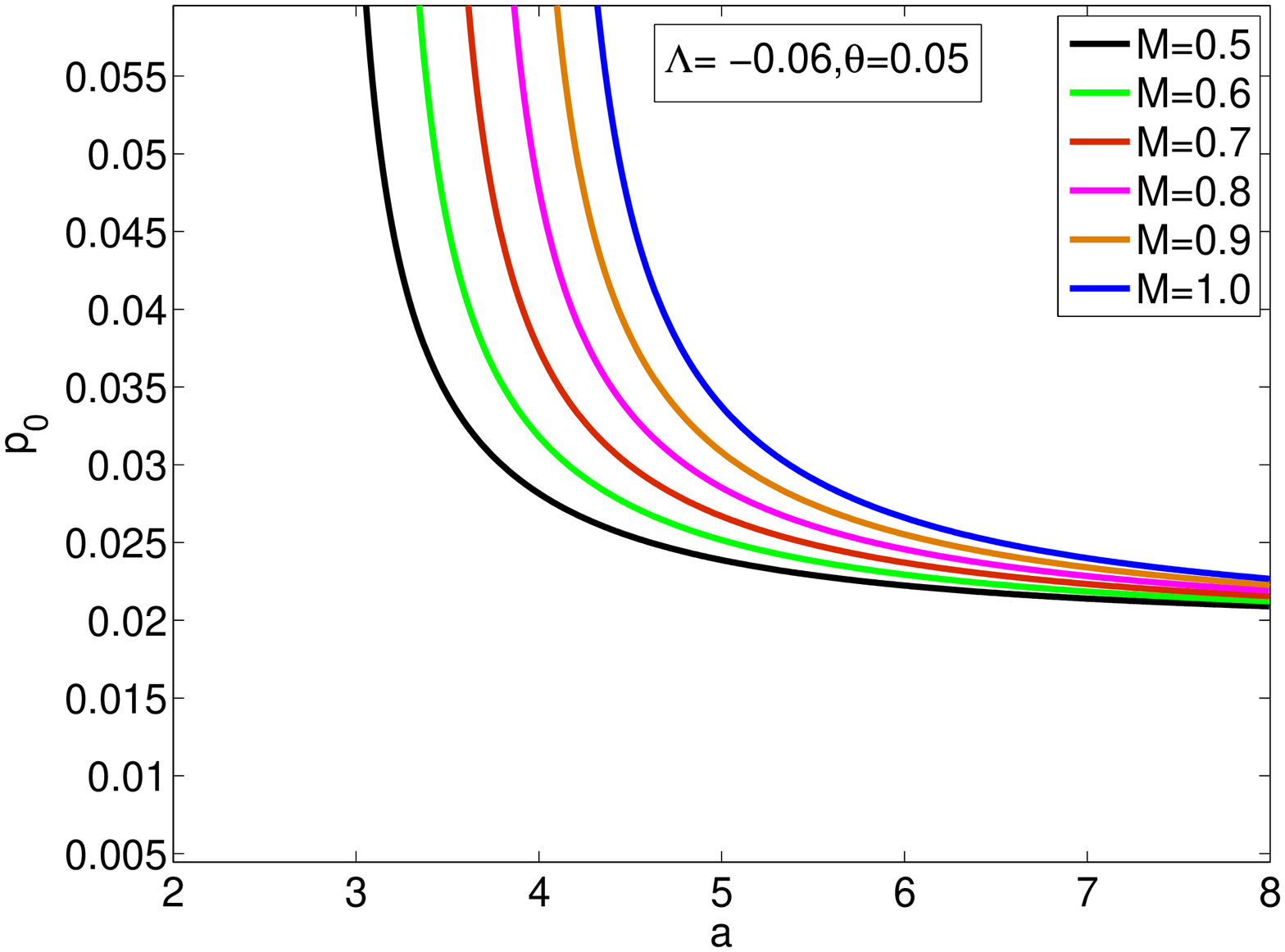} \\
\end{tabular}
\caption{Surface Pressure $p$ has been plotted against $a$ for different values of mass $M$ when noncommutative parameter
 $\theta=0.01$ (left) and for $\theta=0.05$ (right). }
\end{figure*}

\begin{figure*}[thbp]
\begin{tabular}{rl}
\includegraphics[width=10cm]{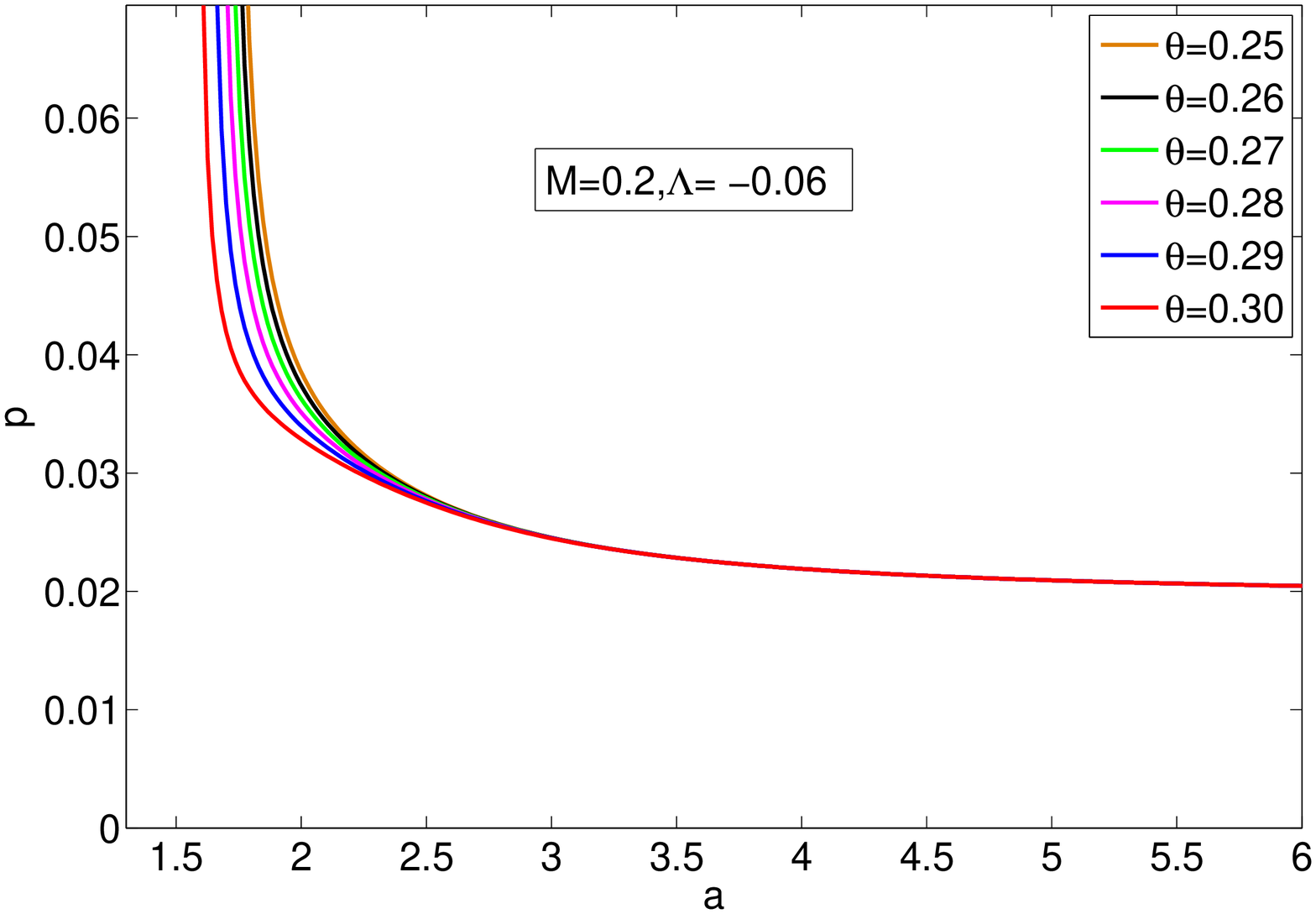}&
\includegraphics[width=12cm]{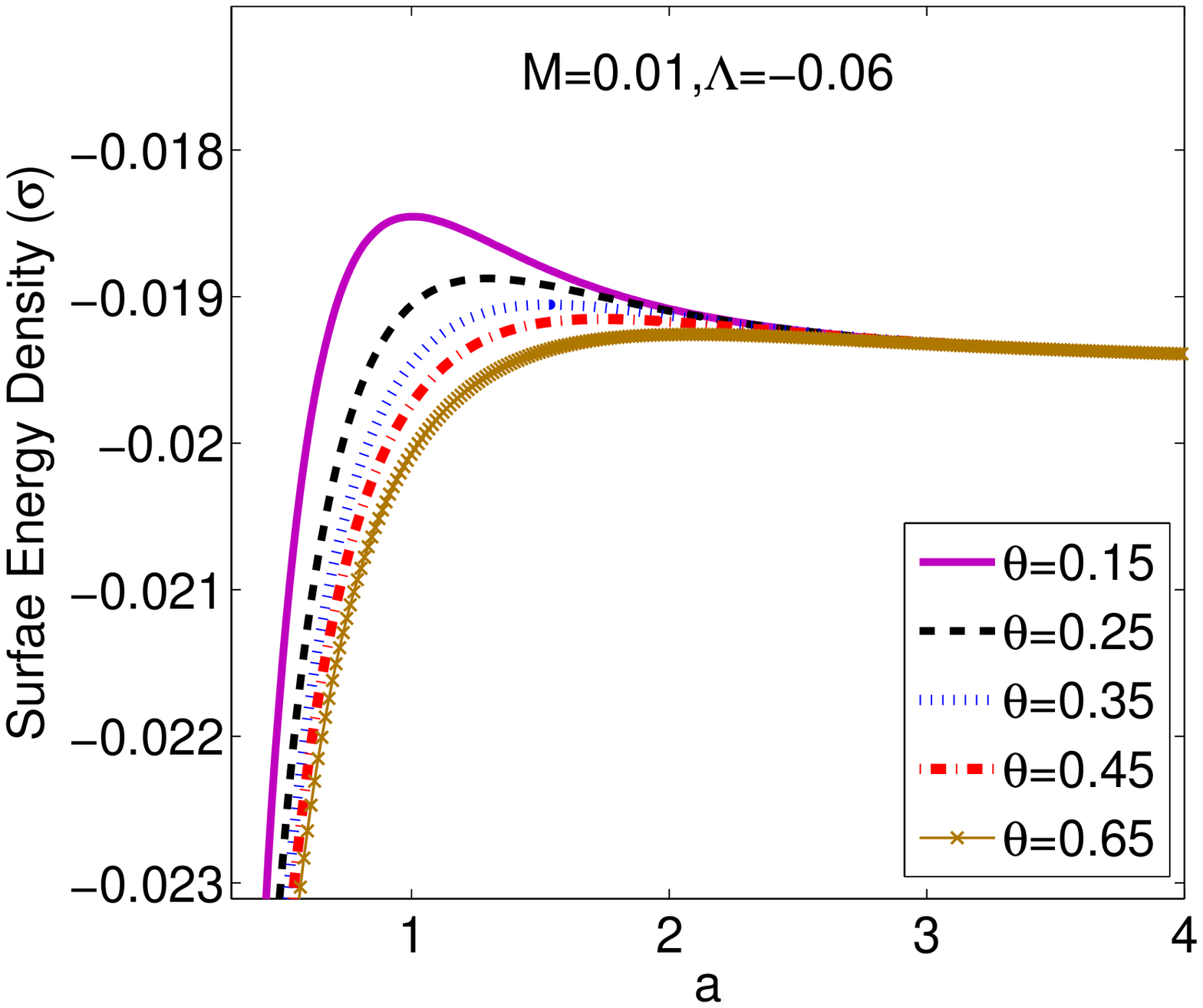} \\
\end{tabular}
\caption{Surface Pressure $p$ has been plotted against $a$ for different values of noncommutative parameter $\theta$ when mass (M) is fixed (left) and surface energy density $\sigma$ has been plotted against $a$ for different values of noncommutative parameter $\theta$ when mass (M) is fixed (right). }
\end{figure*}

\section{The Gravitational Field}
In this section we analyze  the attractive or repulsive nature of the wormhole. For this analysis
 we calculate the observer's three acceleration $a^{\mu}=u^{\mu}_{~;\nu}u^{\nu}$, where $u^{\nu}$ is given by
\begin{equation}
u^{\nu}=\frac{dx^{\nu}}{d\tau}=\left(\frac{1}{\sqrt{f(r)}},0,0 \right).
\end{equation}
The only non-zero component is given by
\begin{equation}
a^{r}=\Gamma_{tt}^{r}=\frac{1}{2}f'(r)=-\frac{Mr}{2\theta}e^{-\frac{r^{2}}{4\theta}}-\Lambda r,
\end{equation}

\begin{figure*}[thbp]
\begin{tabular}{rl}
\includegraphics[width=6cm]{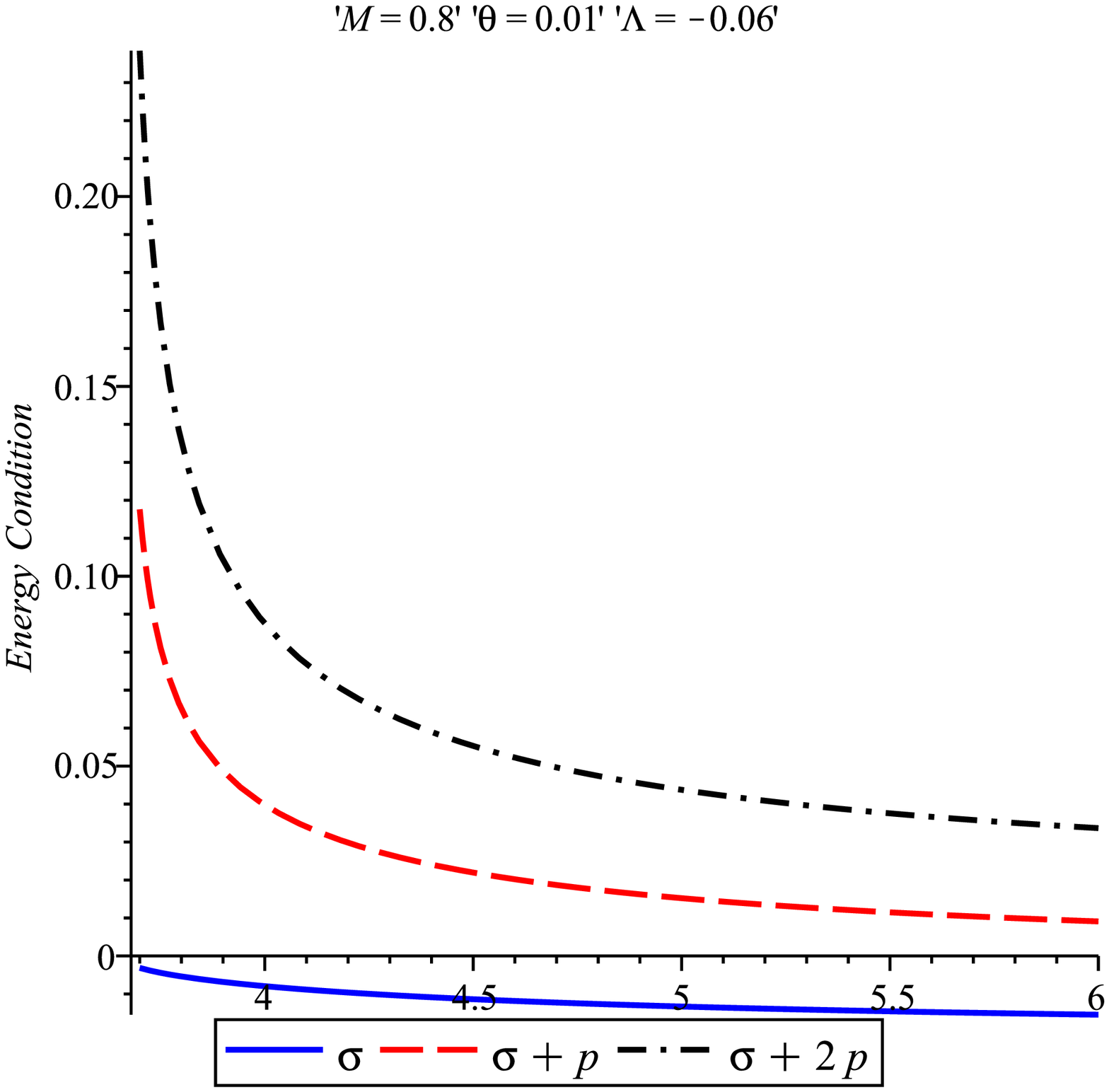}&
\includegraphics[width=6cm]{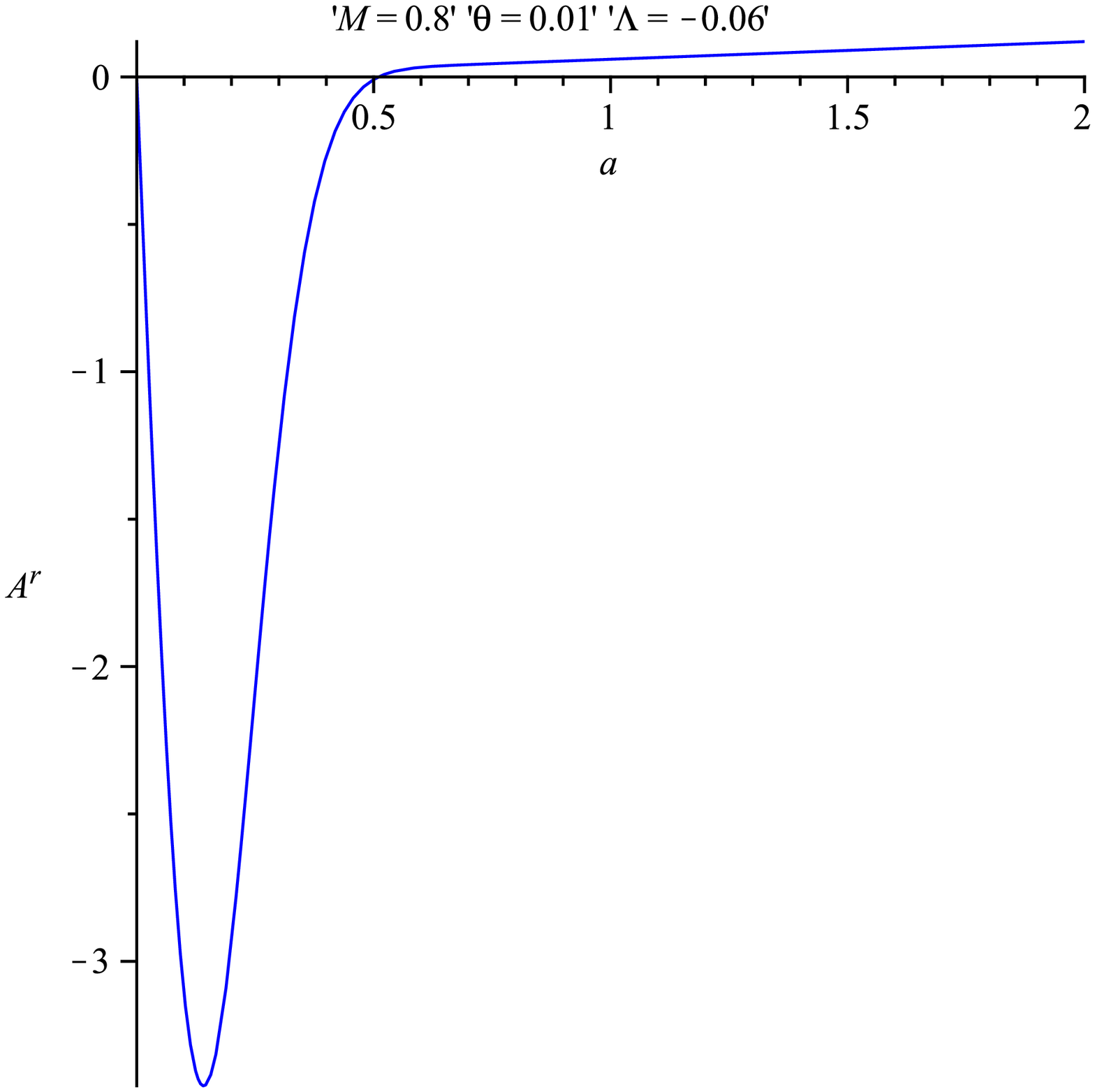} \\
\end{tabular}
\caption{Energy condition is plotted against `a' (left) and $a^{r}$ is plotted against `a'(right).}
\end{figure*}

 A  test particle when radially moving and  initially at rest,
obeys the equation of motion
\begin{equation}
\frac{d^{2}r}{d\tau^{2}}=-A^{r}=\frac{Mr}{2\theta}e^{-\frac{r^{2}}{4\theta}}+\Lambda r.
\end{equation}
If $A^{r}=0$, we get the geodesic equation and solving the eq. (16), for r  we obtain
\begin{equation}
r=\pm \sqrt{-4\theta\ln\left(-\frac{2\theta \Lambda}{M}\right)}.
\end{equation}
Here we consider only positive sign to find the value of `r' since it was previously assumed $a>r_{h}$. \\
Now, the wormhole will be attractive in nature if $A^{r}>0$ i.e., `r' satisfying  the following condition
\begin{equation}
\sqrt{-4\theta\ln\left(-\frac{2\theta \Lambda}{M}\right)}<r<\infty,
\end{equation}
and similarly, the wormhole will be repulsive in nature if $A^{r}<0$, which is given by
\begin{equation}
0<r<\sqrt{-4\theta\ln\left(-\frac{2\theta \Lambda}{M}\right)}.
\end{equation}
The Eqs. (18)-(20) will be  valid  only if $\left|-\frac{2\theta \Lambda}{M}\right|<1$.
Fig. \textbf{4} (right) shows that the curve cuts the `a-axis' at the point $ a=\pm 0.5 $ for a fixed value of
M  = 0.8, $\theta$ = 0.01, and $\Lambda$ = -0.06. Hence, the wormhole will be attractive for $0.5<a<\infty$ and repulsive for
$0<a<0.5$.

\section{Total amount of Exotic Matter}

To construct thin-shell wormholes, we determine the total amount of
exotic matter. Though, using noncommutative BTZ black hole in the
thin-shell wormhole construction is that, it is not
asymptotically flat and therefore the wormholes are not
asymptotically flat. Recently Mazharimousavi, Halilsoy and Amirabi \cite{{Mazharimousavi}}
shown  that a non-asymptotically
flat black hole solution provides stable  thin-shell wormholes
which are entirely supported by exotic matter and
quantified by the integral \cite{{KY},{ES},{MC},{FK}}

\begin{equation}
\Omega=\int[\rho+p_r]\sqrt{-g} d^{2}x,
\end{equation}

\begin{figure*}[thbp]
\begin{tabular}{rl}
\includegraphics[width=7cm]{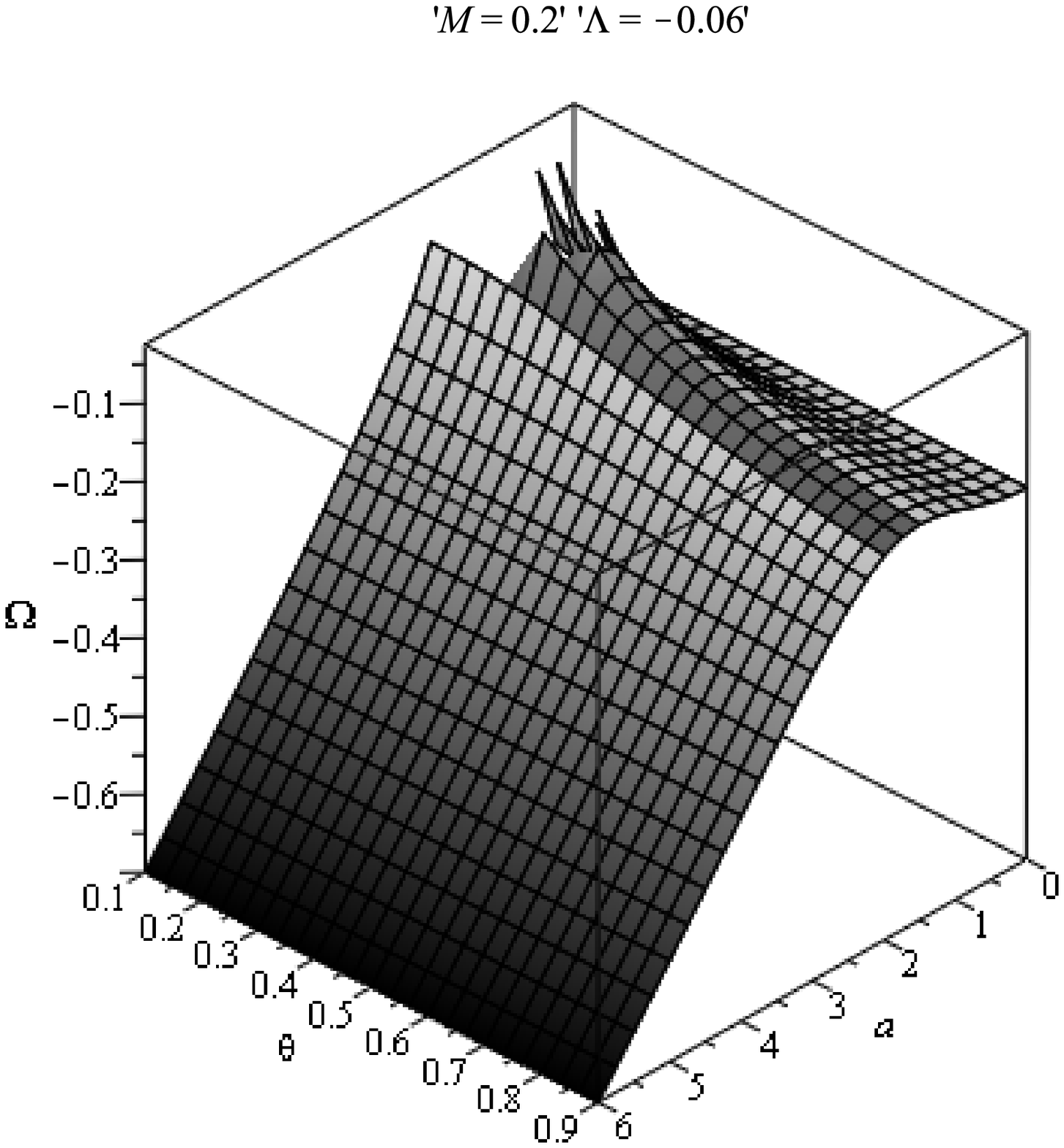}&
\includegraphics[width=7cm]{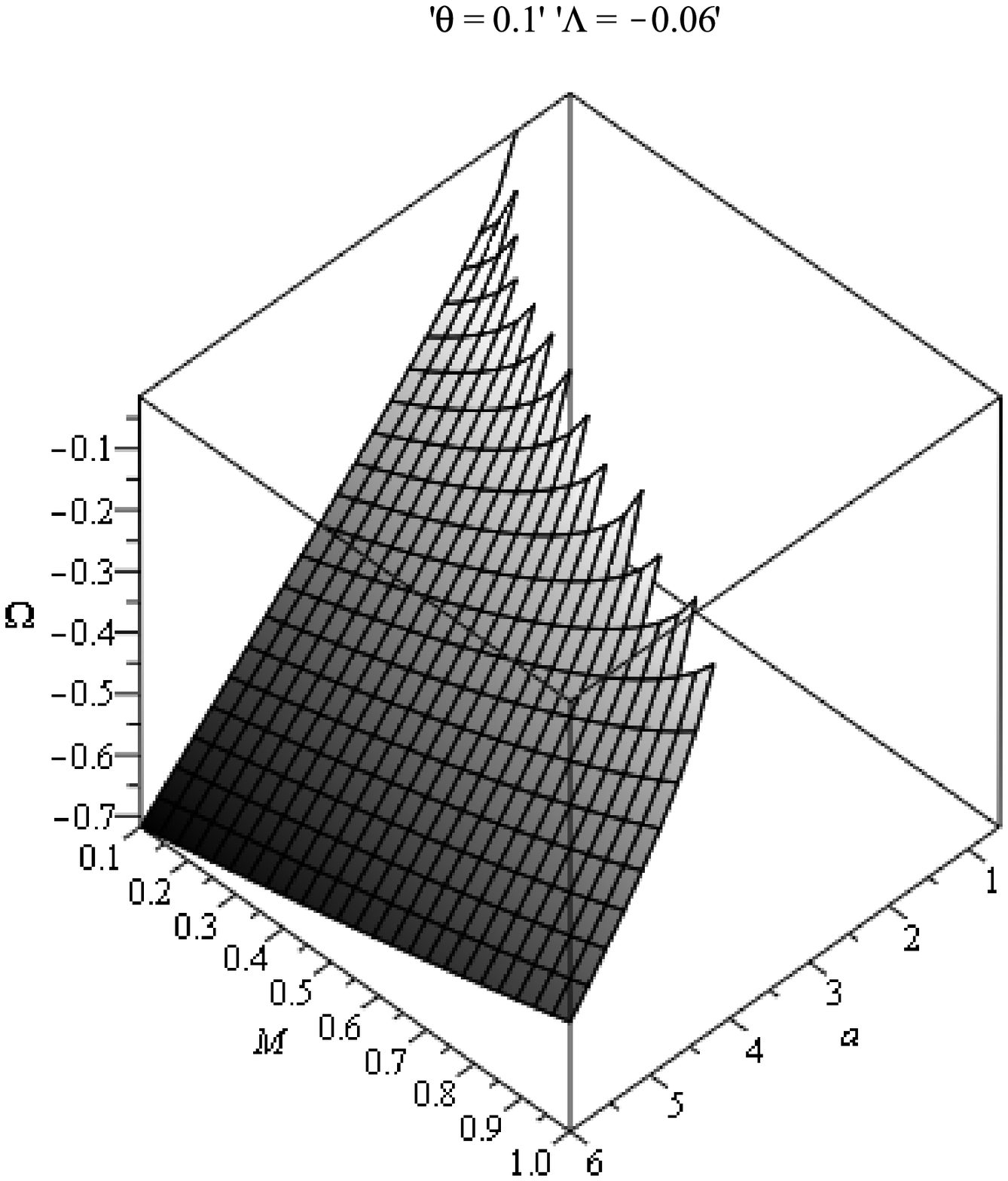} \\
\end{tabular}
\caption{The variation of total amount of exotic matter $\Omega$ has been plotted against $a$ for different values of $\theta$ when mass $M$ is fixed (left) and for different values of mass (M) when noncommutative parameter $\theta$ is fixed (right).}
\end{figure*}

where $g$ represents the determinant of the metric tensor.
Now, by introducing the radial coordinate $R=r-a$, we have
\begin{equation}
\Omega=\int_0^{2\pi}\int_{-\infty}^{\infty}[\rho+p_r]\sqrt{-g}~dR d\phi.
\end{equation}
Since the shell is infinitely thin, it does not exert any radial
pressure, i.e., $p_r=0$ and using $\rho=\delta(R)\sigma(a)$ for the above integral
we have

\begin{equation}
\Omega=\int_0^{2\pi}[\rho \sqrt{g}]|_{r=a}d\phi = 2\pi a \sigma(a)=-\frac{1}{2}\sqrt{M\left(2e^{-\frac{a^{2}}{4\theta}}-1 \right)-\Lambda r^{2}}.
\end{equation}

From Eq. (23), one can see that the total amount of exotic matter depends on the
mass of the black hole and the noncommutative parameter $\theta$. With the help
of graphical representation (see Fig. \textbf{(5)}) we are trying to describe
the variation of the total amount of exotic matter in the shell with respect
to the  mass of the black hole M and the noncommutative parameter $\theta$.
It is clear from Fig. \textbf{(5)}, that we can reduce the total amount of exotic matter for the shell
by increasing the mass M, of the black hole if we fixed the noncommutative parameter $\theta$ or
by choosing very small $\theta$, when the mass M is fixed.

So the mass of the black hole M and the noncommutative parameter $\theta$ plays a
crucial role to reduce the total amount of exotic matter for the proposed thin-shell wormhole.
Furthermore it can be noted that $\Omega\rightarrow 0 $ if $a\rightarrow r_h $ i.e., if $a$ is very
close to the event horizon of the noncommutative BTZ black hole then the required total amount of
exotic matter will be infinitesimal small and if $a>>r_h$ then $\Omega\rightarrow -\infty$.

\section{An Equation of State}
Taking the form of the equation of state (EoS) to be $p=\omega \sigma$, we obtain
from  Eqs. $(13)$ and $(14)$
\begin{equation}
\omega=\frac{p}{\sigma}=\frac{a^{2}\left(\frac{M}{2\theta}e^{-\frac{a^{2}}{4\theta}}+\Lambda \right)}{M\left(2e^{-\frac{a^{2}}{4\theta}}-1 \right)-\Lambda a^{2}}.
\end{equation}
From Eq. $(24)$, we observe that if the location of the wormhole throat is very large
i.e., if $a\rightarrow +\infty$ then $\omega\rightarrow -1$ and if $g(a)=0$ at some $a =a_0$
where $g(a)=\frac{M}{2\theta}e^{-\frac{a^{2}}{4\theta}}+\Lambda$, then $p\rightarrow 0$, but in
that case $a_0< r_h$ (see Fig.{\bf(6)}). So  dust shell is never found.

\begin{figure}[htbp]
    \centering
        \includegraphics[scale=.3]{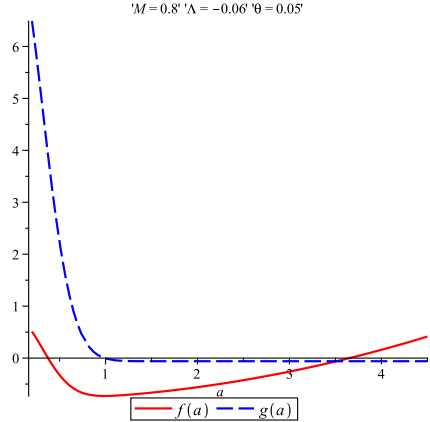}
       \caption{The curve $g(a)$ cuts the 'a' axis at a point $a_0 <r_h $}
    \label{fig:3}
\end{figure}

\section{stability Analysis}
In this section we turn to the question of stability of the wormhole
using two different approaches: (i) assuming a specific
equation of state on the thin shell and (ii) analyzing the stability
to linearized radial perturbations.
\subsection{Dark energy equation of State }
As the negative surface energy density supports the presence of exotic
matter at the throat, we want to study specific cases by giving an equation of state
rather than the analysis for a general equation of
state of the form p= p($\sigma$). For the dynamical characterization of the shell,
we consider the dark energy as exotic matter on the shell which is
governed by an equation of state of the form
\begin{equation}
p=\omega \sigma ~~~~ \text{with}~~~ \omega<0,
\end{equation}
which is a possible candidate for the accelerated expansion of the Universe
and consequently violates the null energy condition.
Such a dark energy fluid can be divided into three cases: a
normal dark energy fluid when -1 $< \omega <$ 0, a cosmological
constant fluid when $\omega = -1$, and a phantom energy
fluid when $\omega <- 1$ . Now, using the Eqs. (25) into (12), we obtain
\begin{equation}
\sigma=\sigma_0\left(\frac{a_0}{a} \right)^{1+\omega},
\end{equation}
 where $a=a_0$ is the static solution and  $\sigma_0=\sigma(a_0)$.
Rearranging Eq. $(9)$, one can obtain
\begin{equation}
\dot{a}^{2}+V(a)=0,
\end{equation}
where the potential V(a) is defined by
\begin{equation}
V(a)=f(a)-16\pi^{2}a^{2}\sigma^{2}.
\end{equation}
\begin{figure*}[thbp]
\begin{tabular}{rl}
\includegraphics[width=7cm]{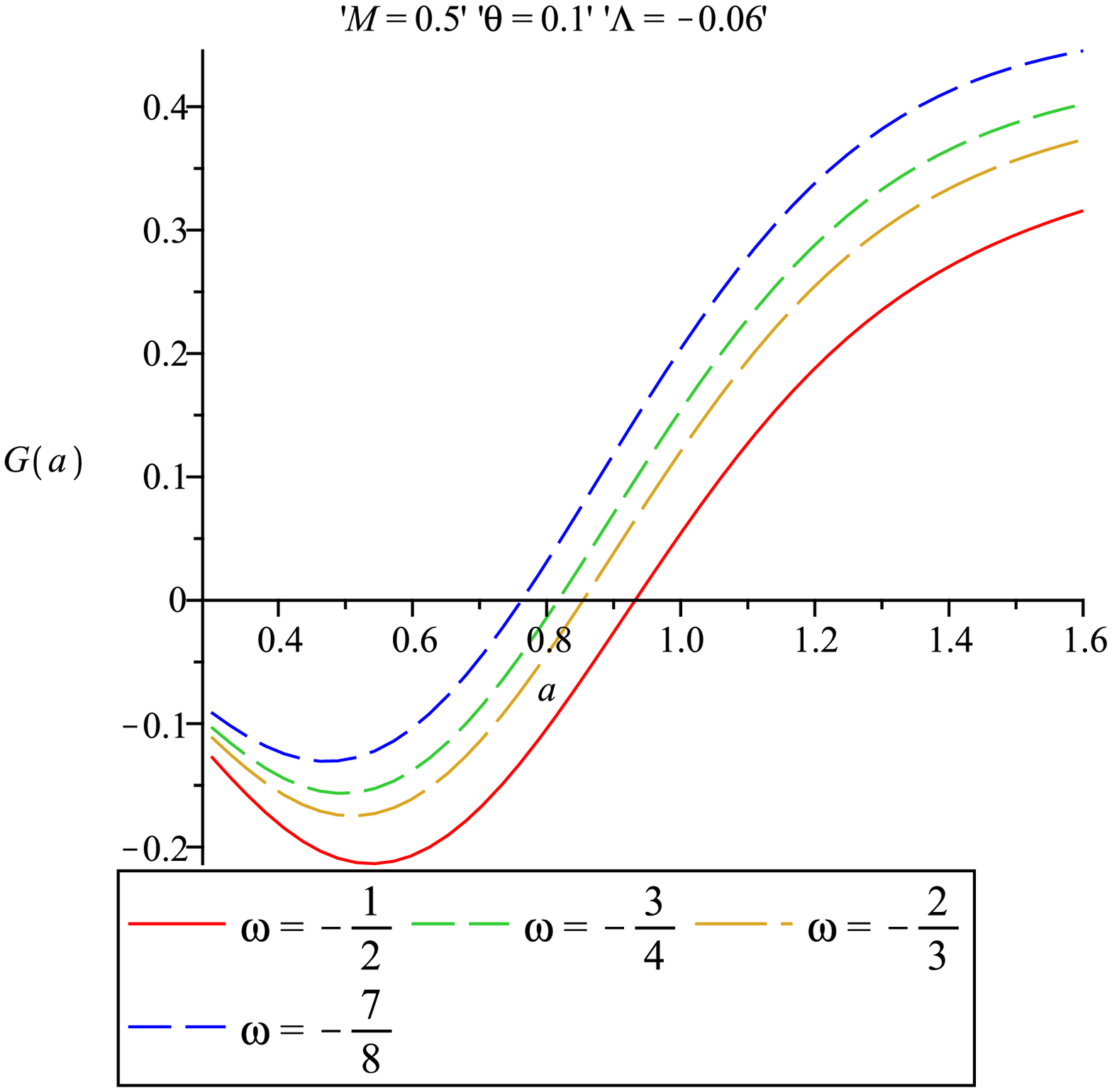}&
\includegraphics[width=7cm]{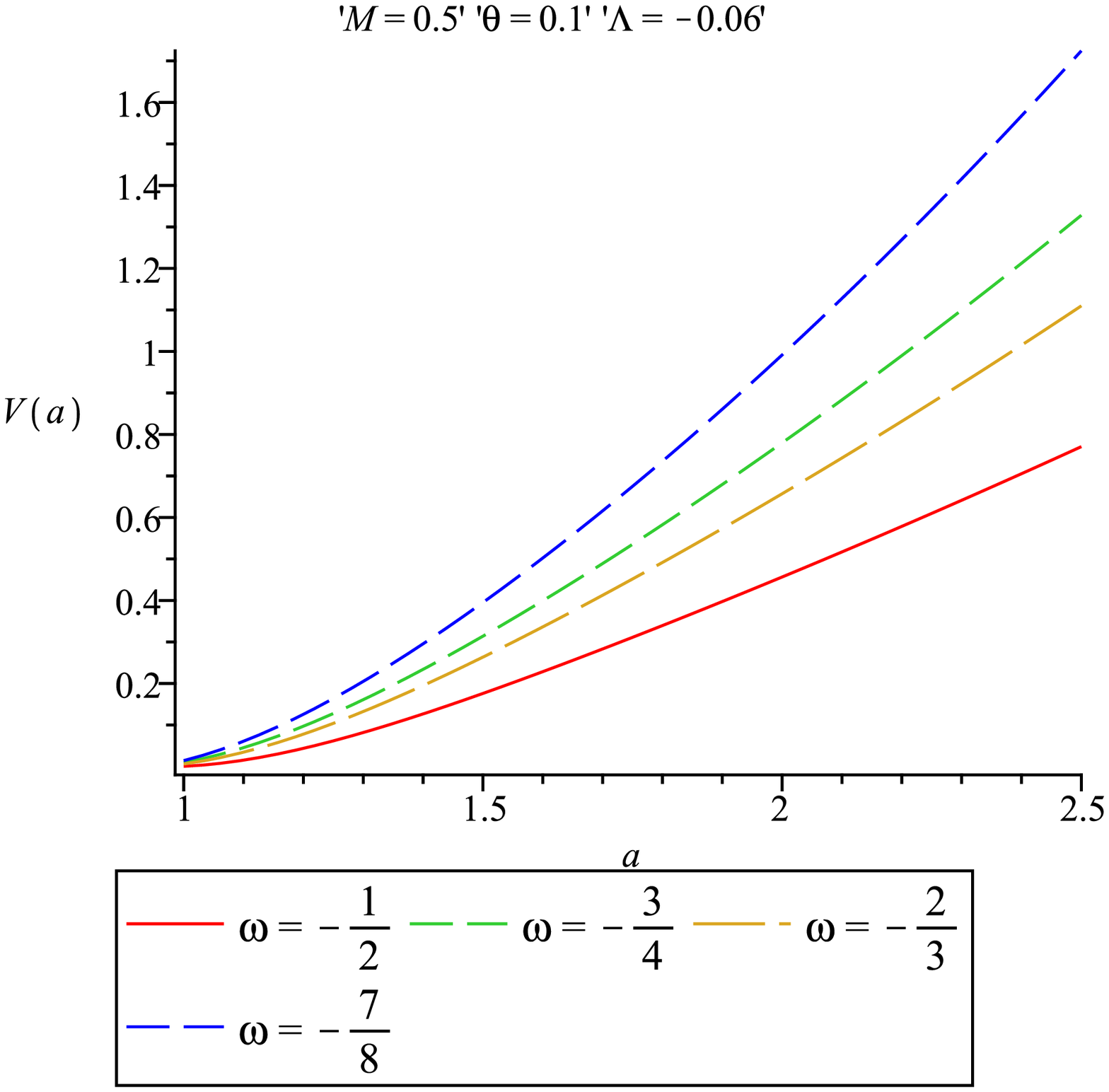} \\
\end{tabular}
\caption{The throat of the thin-shell-wormhole occurs where `G(a)' cuts the `a' axis (Left) and
wormholes collapse for the value of $-1<\omega<-\frac{1}{3}$.}
\end{figure*}

\begin{figure*}[thbp]
\begin{tabular}{rl}
\includegraphics[width=7cm]{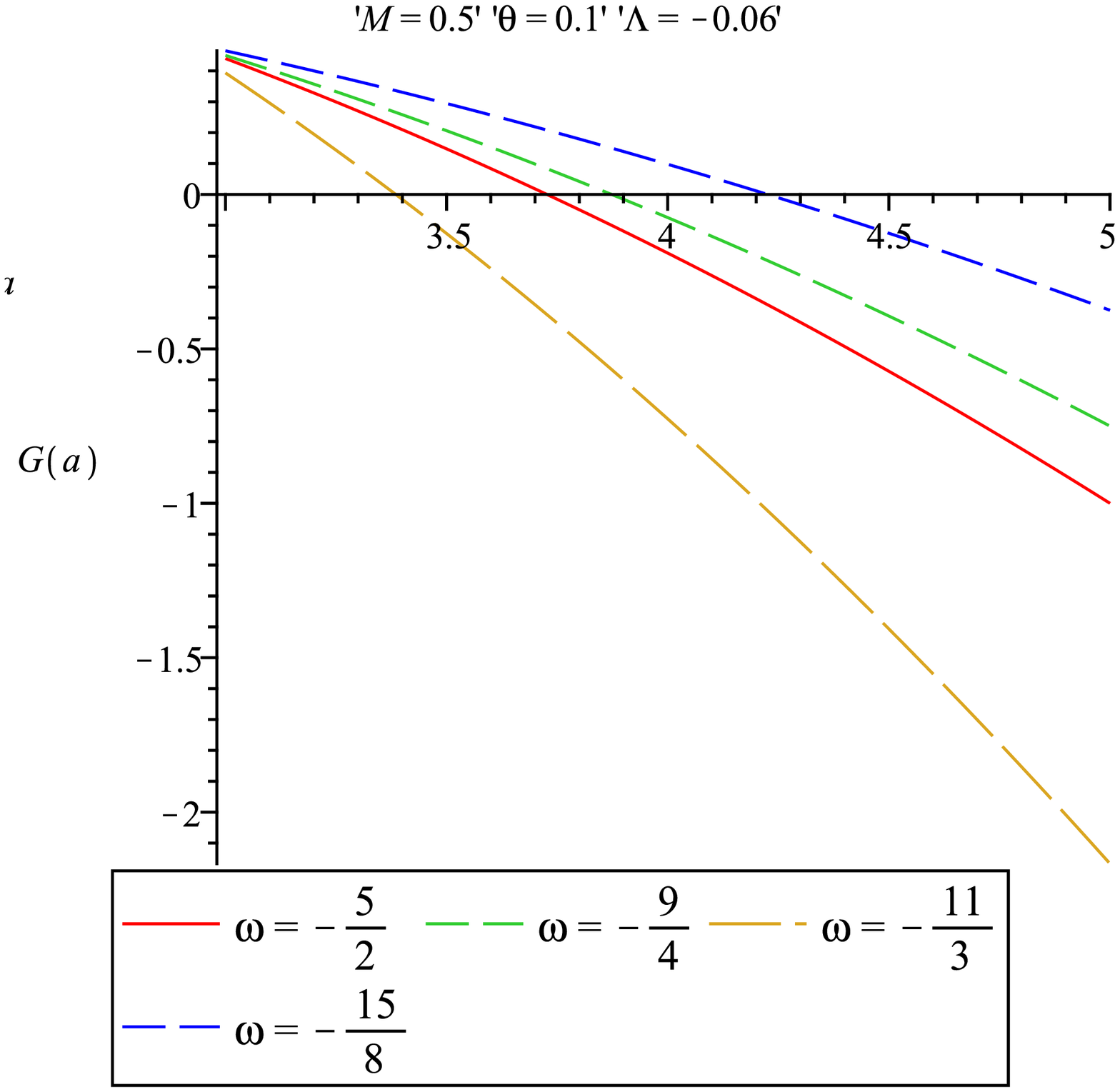}&
\includegraphics[width=7cm]{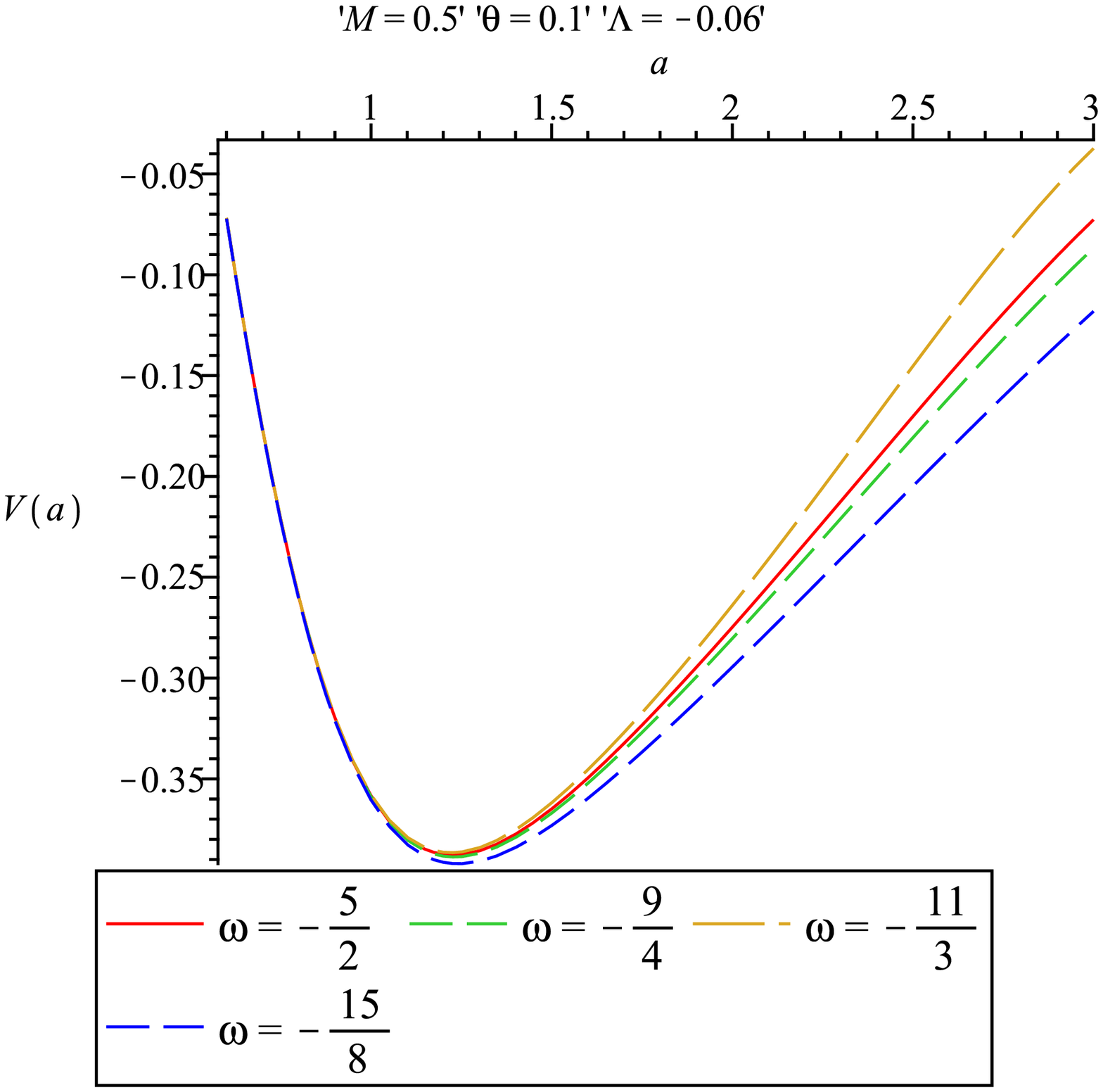} \\
\end{tabular}
\caption{The throat of the thin-shell-wormhole occurs where `G(a)' cuts the `a' axis (Left)
and wormholes are stable for the value of $\omega<-1$.}
\end{figure*}

Now employing the Eq. $(26)$ into $(28)$ and substitute the value of $\sigma(a)$, we get
\begin{equation}
V(a)=M\left[2e^{-\frac{a^{2}}{4\theta}}-1\right]-\Lambda a^{2}-
\left(\frac{a_0}{a} \right)^{2\omega}\left[M\left(2e^{-\frac{a_0^{2}}{4\theta}}-1 \right)-\Lambda a_0^{2}  \right].
\end{equation}
For the linearized stability criteria, we use Taylor expansion to V (a) (upto
second order) around the static solution at $a=a_0$, we obtain
\begin{equation}
V(a)=V(a_0)+(a-a_0)V'(a_0)+\frac{(a-a_0)^{2}}{2}V''(a_0)+O[(a-a_0)^{3}],
\end{equation}
where prime denotes derivative with respect to `a'. The wormhole is stable
 if and only if $V (a_0)$ has local
minimum at $a_0$ and $V''(a_0)> 0$. From Eq. (29) the first derivative of V (a) is
\begin{equation}
V'(a)=-\frac{Ma}{\theta}e^{-\frac{a^{2}}{4\theta}}-2a\Lambda+2\omega a_0^{2\omega}\left[M\left(2e^{-\frac{a_0^{2}}{4\theta}}-1 \right)-\Lambda a_0^{2}  \right] a^{-2\omega-1}.
\end{equation}
The second derivative of the potential is
\begin{equation}
V''(a_0)=-\frac{M}{\theta}\left[1-\frac{a_0^{2}}{2\theta} \right]e^{-\frac{a_0^{2}}{4\theta}}-2\Lambda-\frac{2\omega(2\omega+1)}{a_0^{2}}\left[M\left(2e^{-\frac{a_0^{2}}{4\theta}}-1 \right)-\Lambda a_0^{2}  \right].
\end{equation}

Now the configuration will be stable if and only if $V''(a_0)>0$ for the choice of parameters
M, $\theta$ and $\omega$. We are trying to describe the stability of the
configuration with help of graphical representation due to complexity of the expression. In
Figs. \textbf{7-8}, we plot the graphs to find the possible range
of $a_0$ where V $(a_0)$ possess a local minimum. Using the Eq. (25), we
obtain the throat radius of the shell at some a = $a_0$ when G(a) cuts the
the a-axis which represents the radius of the throat of
shell and using the value of $a_0$ we find that V$(a_0)$ posses a local minima
when $\omega~< -1$ for different values of $\omega$, respectively.
Thus, we can obtain stable thin-shell wormholes supported by exotic
matter filled with phantom energy EOS.

\subsection{Linearized Stability}
We shall study the stability of the configuration under small perturbations
around the static solution i.e., $a=a_0$, preserving the symmetry. Applying the Taylor series
expansion for the potential $V(a)$ upto second order around the stable solution at $a= a_0$,  which provides
\begin{equation}
V(a)=V(a_0)+(a-a_0)V'(a_0)+\frac{(a-a_0)^{2}}{2}V''(a_0)+O[(a-a_0)^{3}],
\end{equation}
where the prime denotes derivative with respect to $d/da$ and $a_0$ is the radius of the static solution.
Since we are linearizing around $a=a_0$, the stability of static solutions requires $V(a_0)=0 $, $V'(a_0)=0$.
Now, the wormhole is stable if and only if $V''(a_0)>0$. To know whether the equilibrium solution is stable or not,
we shall follow the procedure adapted by Poisson and Visser  \cite{Poisson}. The physical interpretation of
$\beta^2$ is a matter of some subtlety which we shall subsequently discuss. First of all we shall restrict ourselves
only to the case when $\beta^2> 0$. The parameter $\beta^2$, when lies in the range of
 $0 < \beta^2 \leq 1$, it can be interpreted as the square of the velocity of sound on the shell
 and the interpretation is not valid when $\beta^2> 1$, because it would mean
a speed greater than the velocity of light, implying the violation of causality. Now
defining the parameter  $\beta^2$ by

\begin{equation}
\beta^{2}=\frac{\partial p}{\partial \sigma}\Bigl\lvert_{\sigma}.
\end{equation}
Since $\sigma^{\prime}={\dot{\sigma}/\dot{a}}$, using the Eq. (12), we have
 $\dot{\sigma}=-\frac{\dot{a}}{a}(\sigma+p)$. The second derivative of the potential is
\begin{equation}
V''(a)=-\frac{M}{\theta}e^{-\frac{a^{2}}{4\theta}}\left(1-\frac{a^{2}}{2\theta}\right)
-2\Lambda-32\pi^{2}p^{2}-32\pi^{2}\beta^{2}\sigma(p+\sigma).
\end{equation}

\begin{figure}[htbp]
    \centering
        \includegraphics[scale=.35]{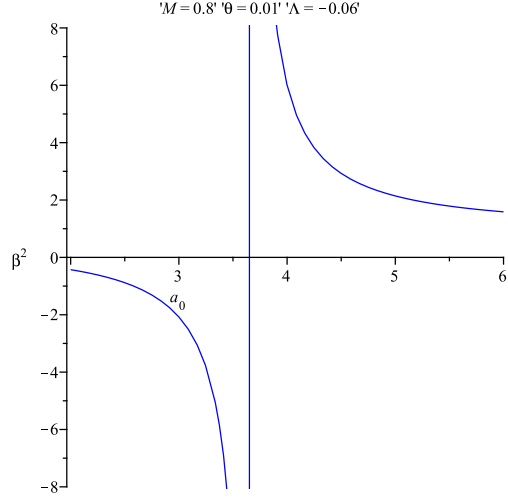}
       \caption{Stability regions for thin-shell wormholes corresponding
to $M=0.8$, $\theta=0.01$ and $\Lambda=-0.06$.}
    \label{fig:3}
\end{figure}

Since we are linearizing around $ a = a_0 $, we require that $ V(a_0) = 0 $ and
$ V^{\prime}(a_0)= 0 $.

Now the configuration will be stable if  $V''(a_0)>0$, which gives
\begin{equation}
\beta^{2}<\frac{-\frac{M}{\theta}e^{-\frac{a^{2}}{4\theta}}(1-\frac{a^{2}}{2\theta})
-2\Lambda-32\pi^{2}p^{2}}{32\pi^{2}\sigma(\sigma+p)}~~~~~~~if~(\sigma+p)<0,
\end{equation}
and
\begin{equation}
\beta^{2}>\frac{-\frac{M}{\theta}e^{-\frac{a^{2}}{4\theta}}(1-\frac{a^{2}}{2\theta})
-2\Lambda-32\pi^{2}p^{2}}{32\pi^{2}\sigma(\sigma+p)}~~~~~~~~~if ~(\sigma+p)>0,
\end{equation}

 Now using the values of $\sigma$ and $p$ from Eqs. (13) and (14), and
solve for $\beta^2 $ by letting $ V^{\prime\prime}(a_0)= 0 $, we have
\begin{equation}
\beta^{2}=\frac{1}{\frac{2M}{a_0^{2}}\left[\frac{a_0^{2}}{2\theta}e^{-\frac{a_0^{2}}{4\theta}}
+2e^{-\frac{a_0^{2}}{4\theta}-1}  \right]}\left[-\frac{M}{\theta}e^{-\frac{a_0^{2}}{4\theta}}\left(1-\frac{a_0^{2}}{2\theta}\right)
-2\Lambda-2\frac{\left(-\Lambda a_0-\frac{Ma_0}{2\theta}e^{-\frac{a_0^{2}}{4\theta}}\right)^{2}}
{M\left(2e^{-\frac{a_0^{2}}{4\theta}}-1 \right)-\Lambda a_0^{2}}\right].
\end{equation}

 Since $\beta$ represents the velocity of the
sound so it is expected that $0<\beta^{2}\leq 1$ . The profile of $\beta^{2}$ is shown in Fig. \textbf{9},
for fixed values of the parameters $M=0.8$, $\theta=0.01$ and $\Lambda=-0.06$. Since Fig. {\bf (4)} shows that $\sigma+p>0$, so for our model the stability region is given by equation {\bf (37)} i.e., our model is stable above the curve on the right side of the asymptotes and below the curve of the left side of the asymptotes of Fig. {\bf (9)}. In this case we have $\beta^{2}>1$. As we are
dealing with exotic matter, we relaxed the range of $\beta^{2}$ in our stability analysis for
the thin-shell wormholes.

\subsection{SUMMARY AND DISCUSSION}
In this work, we have taken a noncommutative BTZ black hole solution of the (2+1)-dimensional
gravity and followed the cut and paste method for removing the singular part of
this manifold in order to construct  thin-shell wormholes
under the assumption that the equations of state on the shell which defines the throat
have the same form as in the static case when it is perturbed preserving the symmetry.
Though, a disadvantage with using noncommutative BTZ black holes in the thin-shell
wormholes construction is that it is not asymptotically
flat and therefore the wormholes are not asymptotically flat. Therefore, there is some 'matter' at infinity.
The construction allows a graphical description
of both $\sigma$ and p as functions of the radius a of the thin shell,
using various values of the mass M and the noncommutative parameter $\theta$.
Also we get from Fig. \textbf{4} (left) that $\sigma$+p $>0$ and $\sigma$+2p $>0$ are
positive which shows matter contained by the shell violates the WEC but satisfy the NEC
and SEC. Using the same parameters we determine whether the wormhole is attractive
or repulsive. Finally, the total amount of exotic matter
required is determined both analytically and graphically with respect to  mass  and noncommutative parameter.

We draw our main attention on the stability of the shell: It is observed that the matter distribution
 of the shell is phantom energy type $p=\omega\sigma$ with $\omega<0$, and the stability of the
static configurations under radial perturbations was analyzed using the standard potential method
and shown that stable wormholes found for $\omega<-1$ in Fig. \textbf{(8)}.
Next, we have addressed the issue of stability of static configurations under
radial perturbations with a linearized equation of state around the static solution
i.e., at a=$ a_0$. The stability analysis concentrated on the parameter
$\beta^2$, ranges of the parameters for which the stability regions include
is  $0< \beta^2 <1$,  interpreted as the speed of sound on the shell.
We found that stable solutions exist if we relax the range of the
parameter $\beta^2$, which is not clear to us for exotic matter.

\subsection{Acknowledgments}
The authors are thankful to the anonymous referees for their valuable comments to improve the manuscript.  AB wishes to thank the authorities
of the Inter-University Centre for Astronomy and Astrophysics,
Pune, India, for providing Visiting fellowship
under which a part of this work was carried out.

\end{document}